\documentclass[12pt]{iopart} 

\usepackage{epsfig}
\usepackage{graphics}
\usepackage{amssymb}

\renewcommand{\H} {{\cal H}}
\newcommand{\Med} {{\rm median}}
\newcommand{\oned} {{1dLR }}

\begin{document}

\title{Numerical study of the dynamics of some long range spin glass
  models} \author{Alain Billoire} \address{ Institut de physique
  th\'{e}orique, CEA Saclay and CNRS, 91191 Gif-sur-Yvette, France}
\date{ \today}
\begin{abstract}
We present results of a Monte Carlo study of the equilibrium dynamics
of the one dimensional long-range Ising spin glass model. By tuning a
parameter $\sigma$, this model interpolates between the mean field
Sherrington-Kirkpatrick model and a proxy of the finite dimensional
Edward-Anderson model. Activated scaling fits for the behavior of the
relaxation time $\tau$ as a function of the number of spins $N$
(Namely $\ln(\tau)\propto N^{\psi}$) give values of $\psi$ that are not
stable against inclusion of subleading corrections.  Critical scaling
($\tau\propto N^{\rho}$) gives more stable fits, at least in the non
mean field region.  We also present results on the scaling of the time
decay of the critical remanent magnetization of the
Sherrington-Kirkpatrick model, a case where the simulation can be done
with quite large systems and that shows the difficulties in obtaining
precise values for dynamical exponents in spin glass models.
\end{abstract}

\pacs{75.50.Lk, 75.10.Nr, 75.40.Gb}
\date{\today}

\maketitle

\section{Introduction}

The low temperature phase of spin glasses is characterized by an
extremely slow dynamics.  A popular method to study this dynamics
consists of Monte Carlo simulations of small systems, followed by a
finite size scaling analysis, leading to the determination of scaling
exponents. In this paper we will show the difficulties in obtaining
precise values for these dynamical exponents with two examples.

The first problem is the equilibrium dynamics of the one-dimensional
long-range Ising spin glass (\oned\hspace{ -0.2em}) model.  The
one-dimensional long-range Ising spin glass model~\cite{KOT} with $N$
sites is a generalization of the Sherrington-Kirkpatrick spin-glass
(SK) model~\cite{SK}.  The spins ($\sigma_i = \pm 1$ with equal
probability) are placed equidistantly on a circle of circumference
$N$, and the Hamiltonian is given by the expression

\begin{equation}
\H = -\sum_{1\leq i < j \leq N} \sigma_i \sigma_j J_{i,j}\ ,
\end{equation}
where the $J_{i,j}$ are independent quenched random couplings
distributed according to

\begin{equation}
J_{i,j}=c_N(\sigma) \epsilon_{i,j} /r_{i,j}\ , \qquad r_{i,j}=N/\pi 
\sin((i-j)\pi/ N)\ ,
\end{equation}
where the $\epsilon_{i,j}$ are iid random variables with zero mean and
unit variance (namely $E(\epsilon_{i,j})=0$ and
$E(\epsilon_{i,j}^2)=1$). The constant $c_N$ is determined by the
normalization condition $\sum_{j\neq 1} E(J_{1,j}^2)=1$.  Depending on
the value of the real parameter $\sigma\geq 0 $, the model behaves as
follows:

\begin{itemize}
\item $0\leq \sigma < 1/2$. The static properties of the model do not
  depend on $\sigma$~\cite{MORI,FULL5} and are thus the same as for
  the SK model (obtained for $\sigma=0$). In particular the critical
  temperature is $T_c=1$ independent of $\sigma$. This is the SK like
  region;
\item $1/2 \leq \sigma \leq 2/3$. The model is in a mean field phase
  below some $T_c>0$;
\item $2/3 \leq \sigma \leq 1$. The model is in a non mean field phase
  below some $T_c>0$;
\item $1 < \sigma$. The model has no phase transition, namely $T_c=0$.
\end{itemize}

This model has been extensively studied numerically (both in this
version~\cite{FULL1}-\cite{FULL4} and in dilute
versions~\cite{DILUTE,VICTOR} that are more suitable for efficient
numerical simulations) since its behavior as a function of $\sigma$ is
analogous to the behavior of the Edwards Anderson Ising (EAI) spin
glass model (the nature of the low temperature of this model has been
the subject of a controversy for many years) as a function of the
dimension $d$: in the \oned model the value $\sigma=2/3$ plays the
role of the upper critical dimension of the EAI model ($d_{ucd}=6$ for
this model) and the value $\sigma=1$ plays the role of the lower
critical dimension ($d_{lcd}=2$ for this model). The hope has been
expressed that there is an exact correspondence between $\sigma$ and
$d$ and that there is in particular two values of $\sigma$ defining
\oned models with the same universal properties as the EAI model in
dimensions $4$ and $3$ respectively.  This hope is not substantiated
by recent numerical results (using the diluted version of the model)
however~\cite{VICTOR}.

In a recent paper, Monthus and Garel~\cite{MONGAR1D} give numerical
evidence that the equilibrium relaxation time $\tau$ of the \oned
model with Gaussian distributed quenched couplings scales like
$E(\ln(\tau))\propto N^{\psi}$ with $\psi\approx 1/3$, like in the SK
model (see~\cite{ABEM,AB1,AB2,MONGAR} and references therein), for all
values of $\sigma \in[0,1]$. This is a surprising result since the
static properties of the model do depend on $\sigma$ for $\sigma>
2/3$, and it contradicts the later analytical prediction
that~\cite{MONT} $\psi=1-\sigma$ in this region.  The results
of~\cite{MONGAR1D} have been obtained for the temperature $T=T_c/2$,
but it is natural to expect that $\psi$ is temperature independent
below $T_c$.  In~~\cite{MONGAR1D} the relaxation time is defined from
the long time exponential decay of the equilibrium spin spin
autocorrelation function $q_J(t)$ and is obtained numerically, for a
given disorder sample $J$, using an eigenvalue technique introduced
in~\cite{MONGAR}.  The systems studied in~\cite{MONGAR1D} are however
very small with $6 \leq N \leq 20$.  Our purpose is to investigate the
question further by direct Monte Carlo simulation of the dynamics of
the model, a method that allows using quite larger system sizes. It is
however subject to thermal errors and implies the use of the median to
analyze the data, instead of the average, due to the presence of a
tail of very slow disorder samples, whose relaxation time cannot be
measured practically with our method.

The second problem is the critical relaxation of the SK
model. This is a particularly interesting phenomenon from a numerical
point of view since the scaling exponents are exactly known and, as
there is neither the need to equilibrate the system nor to sample
thoroughly the phase space, we can simulate quite large systems, by
spin glass standards at least.

If a very strong constant magnetic field is applied to a spin glass,
the individual spins align along this magnetic field.  If afterward
this magnetic field is switched off, the spin glass relaxes very
slowly towards a state of non-zero remanent magnetization, with some
excess internal energy relative to the internal energy at
equilibrium. This phenomenon has been the object of detailed
experimental studies~\cite{Exp}.

At the critical point this phenomenon is simpler and well understood
analytically~\cite{PaRaRiRu,CaPaRi} for the SK model: Starting at time
$t=0$ from a configuration where all spins are aligned, both the
magnetization $m(t)$, the overlap between two clones $q(t)$ and the
internal energy $e(t)$ relax algebraically towards their equilibrium
values with simple exponents, namely $m(t)\propto t^{-\delta_m}$,
$q(t)\propto t^{-\delta_q}$, $e(t)-e(\infty)\propto t^{-\delta_e}$,
with $\delta_e=\delta_q=1$ and $\delta_m=5/4$.  It has been argued
that below $T_c$, $m(t)$, $q(t$) and $e(t)$ relax in two steps, first
algebraically towards a non equilibrium $N$ dependent value and then,
on some exponentially large time scale (out of reach of numerical
simulations), towards their equilibrium value~\cite{MaPaRo,PaRaRiRu},
however the numerical situation is not clear~\cite{BiPaRi}.

The outlook of this paper is as follows: in a first section we study
the equilibrium dynamics of the \oned model.  The precise
determination of the scaling law  governing the growth with $N$ of
the relaxation time of this model turns out to be quite difficult. In
a second section we turn to another problem, namely the decay 
of the critical remanent magnetization of the SK model. 
This is an enlightening case where the
exact values of the scaling exponents are known and 
we can test the precision of the Monte Carlo method.
 In a final section we
present our conclusions.

\section{Equilibrium dynamics of the \oned model}

We measure the value of the spin spin autocorrelation function
$q_J(t)$ at equilibrium, and define the relaxation time $\tau_J$ as
the (unique) solution of the equation $q_J(\tau_J)\equiv 1/2
\sqrt{<q^2>_J}$, where the average overlap squared is measured in the
same disorder configuration. We have shown~\cite{AB2} that the
alternative definition involving the disorder averaged $<q^2>$, namely
$q_J(\tau_J)\equiv 1/2 \sqrt{E(<q^2>_J)}$ gives very similar results
for the system sizes considered.  As discussed in~ı\cite{AB1} the
method we are using assumes that $q(t)/\sqrt{<q^2>}\approx G(t/\tau)$
with some function $G(\cdot)$.

We now proceed to give some technical details of our simulation: The
autocorrelation function $q_J(t)$ is measured for integer values of
the argument inside a time window of size $W_{T,N}=6000\ R_{T,N}$, where
$R_{T,N}$ is an integer scale factor adjusted in such a way that the
window width scales roughly like $\tau$ when the temperature $T$ and
the number of spins $N$ vary, specifically we enforce the relation
$W_{T,N} \gtrapprox 10\ \Med(\tau)$. For small values of $\tau$
however, $R_{T,N}$ sticks to its lowest possible value $R_{T,N}=1$,
and the bound that we impose is definitively not saturated.  Inside
this time window $q_J(t)$ is measured for $60$ values of the argument
(that are multiple of $R_{T,N}$) whose logarithms are roughly
uniformly distributed. In our computer program the relaxation time
$\tau_J$ is defined as the smallest integer $t_{up}$ multiple of
$R_{T,N}$ such that $q_J(t_{up})< 1/2\sqrt{<q^2>_J}$. As a check of
our procedure we have also measured $t_{int}$, the result of a linear
interpolation between $t_{down}\equiv t_{up}-R_{T,N}$ and $t_{up}$,
truncating the result to the lowest integer, namely in C language
style notations:

\begin{equation}
t_{int} \equiv (int) \Bigl[t_{down}+\frac{q_J(t_{down})
    (t_{up}-t_{down}) }{q_J(t_{down})-q_J(t_{up}) }\Bigr]\ .
\end{equation}
The difference between $t_{up}$ and $t_{int}$ gives an idea of the
systematic errors induced by our selection of values of $t$ for the
measurement of $q_J(t)$. Needless to say we must make such a selection
and cannot compute and store $q_J(t)$ for all values of $t$ inside the
chosen window. We note that in contrast to our approach the method
of~\cite{MONGAR1D} gives real valued relaxation times.

We take the quenched couplings to be Gaussian distributed, like
in~\cite{MONGAR1D} (but in the $\sigma=0$ case, where we consider
binary distributed couplings, in order to compare
with~\cite{AB1,AB2}), and consider $N_{dis}$ independent disorder
samples, with $N_{dis}=1024$ in most cases.  We first bring the system
to equilibrium, using the parallel tempering algorithm, that is
currently the best existing algorithm for this purpose. The lowest
temperature is $T_c/2$ and we have $N_T=16$ values of $T$ separated by
a fixed interval $\Delta T=T_c/10$, namely the highest temperature is
$T=2 T_c$, well inside the paramagnetic phase where the dynamics is
fast. We perform $2 \ 10^5$ parallel tempering sweeps (a parallel
tempering sweep consists of $N_T-1$ conditional swaps of configurations
plus one Metropolis system sweep), the second half being used to
measure static quantities like $<q^2>$ that we will use later. For the
largest systems, we have used $31$ values of the temperature separated
by the fixed interval $\Delta T=T_c/20$, and perform $4 \ 10^5$
parallel tempering sweeps. We have monitored the time $n_T^c$ spent by
a given Markov chain $c$ at each temperature.  With the parallel
tempering algorithm as $N$ increases, at fixed $\Delta T$, one notices
that these $(N_T)^2$ numbers, as measured inside the simulation,
spread more and more around their common mean.  Very soon some $n_T^c$
becomes zero and the algorithm clearly breaks down.  In our
simulation, out of many thousand chains, no chain has a ratio
$r_c\equiv \max_T(\{ n_T^c\})/\min_T(\{ n_T^c\})>64$, and only $3$ a
ration above $32$, a reasonable result according to our experience
with this algorithm, if not ideal.  This equilibrium procedure is done
for two independent copies of the system with the same disorder
samples, two clones.

Next we measure (for each clone) the autocorrelation function $q_J(t)$
along a long chain of length $N_{sweeps}$ of at least $10^7$ Metropolis
sweeps, and always larger than ten window length. Measurements are
made every $5 R_{T,N}$ sweeps. We have checked that a ten fold
increase in the chain length changes very little to the estimated
median and to the estimated statistical error on this median, indeed in
this simulation the main source of statistical error is the
fluctuations of the disorder and not the thermal
noise~\cite{AB1,AB2}. The data for $q_J(t)$ are averaged over the two
clones.

We base our statistical analysis of the dynamics of the model on the
median of the relaxation time distribution. Considering the median
rather than the average leads to an immense saving in computer time:
we only need the value of $\tau$ for at least the $50 \%$ faster
disorder samples (samples with the smaller values of $\tau$), and can
forget about the slower samples (that we take to have $\tau=+\infty$).  Our
empirical choice of window length turns out to be large enough for
this aim.

To compute the median we first sort the $N_{dis}$ data for
$\tau$, and define the median as the average of the values for data
number $N_{dis}/2-1$ and $N_{dis}/2$ (In our simulation the number of
disorder samples $N_{dis}$ is always even). For small system sizes
usually the values of $\tau$ for the data number $N_{dis}/2-1$ and
$N_{dis}/2$ are the same (remember that our $\tau$'s are integer
valued), and the median is accordingly not a very good measure. With
the parameters we have chosen for the time window however this never
occurs for larger systems (that are what matters for our analysis of
the scaling of the relaxation time). The statistical errors on
$\Med(\ln(\tau))$ are obtained from a bootstrap analysis.  We have
done the statistical analysis for both $t_{int}$ and $t_{up}$
prescriptions and found results that differ by quite less than the
estimated statistical error. This check is important as it shows that
our selection of values of $t$ for the measurements does not produce
significant systematic errors. The plots presented here have been made
using $t_{int}$.

In order to compare directly with the results of~\cite{MONGAR1D} we
have performed runs at four values of $\sigma$, namely $\sigma=0$ (SK
model), $\sigma=0.25$ (inside the SK like region), $\sigma=0.75$ and
$\sigma=0.85$ both in the non mean field region. 

\begin{table}[htb]
\begin{center}
\begin{tabular}{|r|rr|rr|r|}
\hline
$\sigma$ & $\mathbb{P}(J)$ & Upgrading& $N_{max}$ &$\ln(\tau)$ \\
\hline
 $0.75$   &  Gauss & Random & 1024& 12.0  \\  
 $0.75$   &  Gauss& Systematic & 512  &  11.1 \\  
 $0.25$   &  Gauss & Systematic & 1024 & 8.5   \\  
 $0$   &  $\pm 1$ & Random & 768  & 8.4 \\  
  $0$   &  $\pm 1$ & Systematic & 768  & 7.4 \\  
 \hline
 
\end{tabular}
\end{center}
\caption {Details of the simulation of the \oned model (with $T=0.7
  T_c$): $\sigma$, coupling probability distribution, type of
  upgrading, largest system size simulated, and the median of the
  relaxation time distribution for $N=512$.  The systematic upgrading
  leads to a faster dynamics.}.
\label{table1}
\end{table}

\begin{figure}[htb]
\centering
\includegraphics*[height=8cm,angle=270]{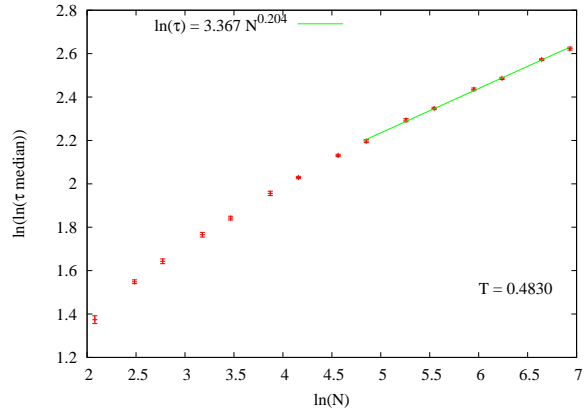} 
\caption{Data for $\ln(\Med(\ln(\tau)))$ as a function of $\ln(N)$ for
  the \oned model with $\sigma=0.75$, $T=0.7\ T_c$ and values of $N$
  between $8$ and $1024$. The quenched random couplings are Gaussian
  distributed and the random updating scheme is used.The green line
  (color on-line) is a two-parameter activated fit of the data for
  $N\ge 128$.}
  \label{fig1}
\end{figure}

Let us start with the \oned $\sigma=0.75$ and $T=0.7$ case. We have
data for systems with $N=8, 16, 24, \ldots , 1024$. We fit these data,
using gnuplot 4.6, to the usual two-parameter form $\Med(\ln(\tau))=C
N^{\psi}$.  We have monitored the evolution of the fitted value of
$\psi$ and of the reduced chi squared $\chi^2/ndf$, as $N_{\min}$ the
lowest value of $N$ included in the fit is increased. A stable plateau
is reached for $\psi$ when $N_{\min}=128$, with a reasonable reduced
chi squared $\approx 0.9$. The value of $\psi$ for such a fit
($128\leq N \leq 1024$) is $\psi=0.20$, with an estimated statistical
error ($\pm 0.003$) on the third digit (But the systematic errors are
larger, as we will see later).  As an example of the stability of the
fit we note that a quite lower cutoff $N_{\min}=32$ would give
$\psi=0.22$ (but an unacceptable $\chi^2/ndf\approx 6$).
Figure~\ref{fig1} shows a plot of $\ln(\Med(\ln(\tau)))$ as a function
of $\ln(N)$ together with the best two-parameter fit. There are
sizable deviations of the data from the fitted curve for low values of
$N$ (and a fit to the low portion of the curve would give a much
larger value of $\psi$ than the value we obtain).  It is natural to
try a fit including some subleading corrections. The data are however
not good enough for a fit to the sum of two power laws, and we have
tried a fit to a power law plus a constant, namely
$\Med(\ln(\tau))=A+C N^{\psi}$.  We find a plateau for $\psi$, with a
reasonable reduced chi squared $\approx 0.3$, for $N_{\min}=32$.  We
will use in the following the same values of $N_{\min}$, namely $128$
and $32$ for fits with two parameters and with three parameters
respectively.  The value of $\psi$ we obtain with this three-parameter
fit is $\psi=0.10\pm 0.01$.  This is many standard deviations away
form the previous result.

We have redone the simulation using systematic upgrading for the
dynamics rather than random upgrading, still using the same
thermalized initial configurations (but with a maximal value of $512$
for $N$).  In the former case the spin configuration at time $t+1$ is
obtained from the spin configuration at time $t$ by applying the
(single spin) algorithm to every sites successively and in a fixed
order In the later case the (single spin) algorithm is applied to a
spin chosen at random, this procedure being repeated exactly $N$
times. It is well known that the former does not satisfy detailed
balance (but do preserve the equilibrium state) whereas the later does
preserve detailed balance. It is usually assumed that both dynamics
lead to the same scaling exponents, due to universality, and we assume
that this is true and compare the results from the two dynamics, in
order to have an idea of systematic errors, obviously not included in
the gnuplot statistical error estimates.  We first remark that the
difference between the random and systematic updating estimates
decreases as $N_{min}$ is increased, as it should. A two-parameter fit
of the systematic dynamics data ($128\leq N \leq 512$) gives $\psi=0.22
\pm 0.1$ and a three-parameter fits ($32\leq N \leq 512$) gives
$\psi=0.08 \pm 0.01$.  Both results are in reasonable agreement with
their random upgrading counterparts, but leave us with some
unexplained discrepancy between the two-parameter and three-parameter
fits.

\begin{figure}[htb]
\centering
\includegraphics*[height=8cm,angle=270]{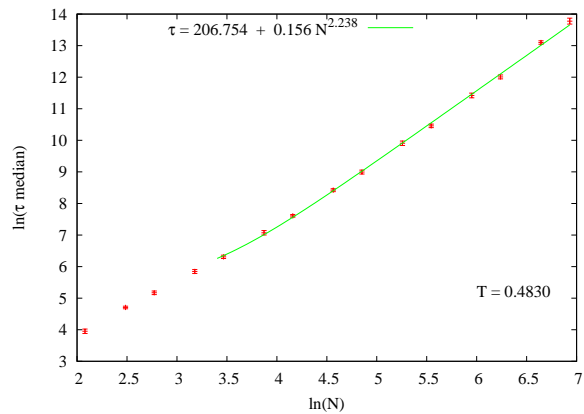}
\caption{Data for $\Med(\ln(\tau))$ as a function of $\ln(N)$ for the
  \oned model with $\sigma=0.75$ and $T=0.7\ T_c$ as in
  Fig.~\ref{fig1}. The green line (color on-line) is a three-parameter
  critical fit of the data for $N\ge 32$.}
  \label{fig3}
\end{figure}

The fact that we obtain very small values of $\psi$ with the
three-parameter fit indicates that may be $\psi$ is indeed exactly
zero, and that we have critical scaling, namely $\tau\propto
N^{\rho}$.   It turns out that critical
scaling fits work nicely. A two-parameter critical fit for random
update gives $\rho=2.29\pm0.05$ with a fair reduced chi squared
($\approx 1.4$) for $N_{\min}=128$, and a three-parameter fit gives
$\rho=2.24\pm0.03$ for $N_{\min}=32$ (see Fig.~\ref{fig3}) with a fair
reduced chi squared ($\approx 1.5$). Systematic upgrade gives
$\rho=2.15\pm0.04$ and $\rho=2.13\pm0.02$ respectively.  There is a
nice overall agreement between the above four estimates of $\rho$,
with systematic errors of few percents.

The values obtained for the exponent
$\rho$ are not absurd, 
it has been argued that the non equilibrium correlation length of the
3d~\cite{z3} and 4d~\cite{z4} EAI model grows like $\xi(t) \propto
t^{1/z(T)}$ where $t$ is the time since the quench, and $z(T)\approx
z(T_c) T_c/T$ with $z(T_c)=6.86 \pm 0.16$ in $3d$ and $z(T_c) \approx
5.4$ in $4d$. That $\xi(t) \propto t^{1/z(T)}$ in an infinite volume
non equilibrium situation means hand-wavily that at time $t$ the
length scales below $\xi(t)$ are equilibrated.  Crudely speaking it
indicates that at equilibrium on a system of size $L$ the relaxation
time $\tau$ fulfills the relation $L\propto \tau^{1/z(T)}$, namely $
\rho=z(T)/d$, not far from the values of the critical scaling fits
of our \oned data (according to~\cite{VICTOR} the \oned model with
$\sigma=0.79$  is a proxy to the 4d EAI model, and $z(T_c) \approx
5.4$ translates into $\rho(T/T_c=0.7)=5.4/(0.7\ d)= 1.9$).

We have made simulations for $T=0.5$ with the same value of $\sigma$
but we can only simulate systems up to $N=192$, where already
$\Med(\ln(\tau))=15.2$, namely $\Med(\tau)\approx 4\thinspace10^6$,
compared to $\Med(\ln(\tau))=13.8$ for $N=1024$ in the $T=0.7$
case. Such a value for $N$ is (jugging from the $T=0.7$ case) quite
too small to obtain sensible results for $\psi$. We also made
preliminary runs for $\sigma=0.85$ with $T=0.7$ and $0.5$ with the
same no-go conclusion.

\begin{figure}[htb]
\centering
\includegraphics*[height=8cm,angle=270]{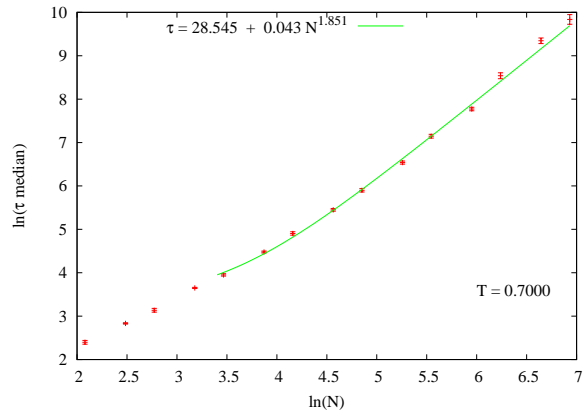}
\caption{Same plot as Fig.~\ref{fig3} but in the SK phase.  The
  critical ansatz fails to reproduce the data. Here $\sigma=0.25$,
  $T=0.7\ T_c$ and values of $N$ are between $8$ and $1024$. The
  quenched random couplings are Gaussian distributed and the random
  updating scheme is used. The green line (color on-line) is a
  three-parameter scaling fit of the data for $N\ge 32$.}
  \label{fig3b}
\end{figure}

Simulations for lower values of $\sigma$ are easier. We have data for
$\sigma=0.25$ and $T=0.7$, namely in a region where the statics is the
same as the one of the SK model. We use random upgrading with systems
up to $N=1024$.  Two-parameter fits give $\psi=0.25\pm 0.01$.
Three-parameter fits give $\psi=0.19\pm 0.02$ (see
Fig.~\ref{fig3b}). In this case there is no satisfactory critical fit:
a two-parameter fit (still with $N\geq 128$) has a reduced chi squared
of 5, and a three-parameter fit ($N\geq 32$) a reduced chi squared of
5 again.

\begin{figure}[htb]
\centering
\includegraphics*[height=8cm,angle=270]{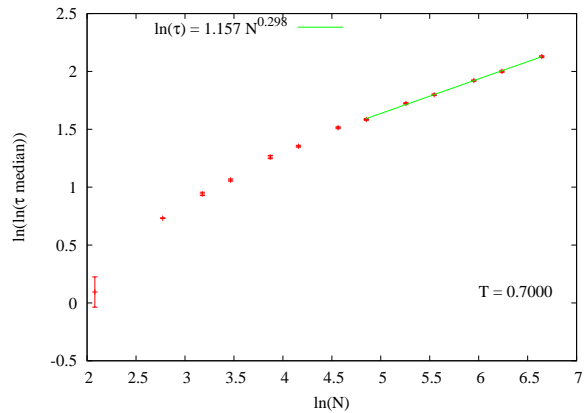} 
\caption{Data for $\ln(\Med(\ln(\tau)))$ as a function of $\ln(N)$ for
  the \oned model with $\sigma=0$ (SK model), $T=0.7\ T_c$ and values
  of $N$ between $8$ and $768$. The quenched random couplings are
  binary distributed and the systematic updating scheme is used. The
  green line (color on-line) is a two-parameter activated fit of the
  data for $N\ge 128$.}
  \label{fig2}
\end{figure}

In order to compare with existing results for the SK model we have
done a simulation of this model with binary distributed
couplings,using systematic upgrading and $N$ up to $768$.  A
two-parameter fit gives $\psi=0.30 \pm 0.01$ (see Fig.~\ref{fig2}), in
agreement with the existing literature~\cite{ABEM,AB1,AB2}. A three
parameter fit gives however a smaller value $\psi=0.17 \pm 0.03$.
Repeating the simulation with random upgrading we obtain $\psi=0.26$
and $\psi=0.17\pm 0.02$ with two-parameter and three-parameter fits
respectively. In this case again, no satisfactory critical fit is obtained.

The conclusion of this section is that our data disagree with the
claim that $\psi= 1/3$ in the spin glass phase for all
$0\leq\sigma<1$. We are left with the following possibilities
i)~Either the activated three-parameter fits are not trustworthy, and
$\psi$ do decrease from a value close to $1/3$ for the SK case to
lower values for larger values of $\sigma$; ii) Or $\psi$ is quite
smaller than previously thought, possibly exactly zero in the non mean
field region; iii) Another possibility is that the coefficient in
front of $N^{\psi}$ is extremely small as found in~\cite{AsBlBrMo} for
the SK model at $T=T_c/20$.  In order to distinguish between theses
scenarios one would need some new ideas, as brute force is not a
possibility here.

\section{The relaxation of the critical SK model}

We simulate the SK model with binary distributed couplings, using the
Heat Bath algorithm with random site updating, starting from a
configuration where all spins are set to one.  The number of sites $N$
ranges from $N=1024$ to $2^{17}=131072$.  We perform $600$ Monte Carlo
time steps, and average the results over $N_{dis}=1310720/N$ disorder
samples.  This scaling of $N_{dis}$ is such that the estimated
statistical errors are roughly $N$ independent.

\begin{figure}[htb]
\centering
\includegraphics*[height=8cm,angle=270]{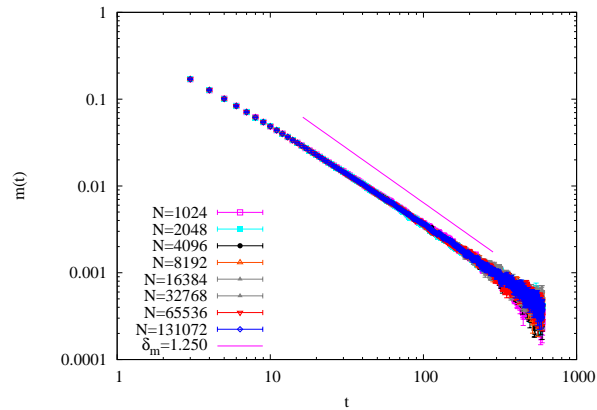}   
\caption{ Decay of the magnetization $m(t)$ as a function of $t$ at
  the critical point for the SK model.  The predicted scaling
  $m(t)\propto t^{-5/4}$ is shown. This is in rough but not perfect
  agreement with the data.}
\label{fig5}
\end{figure}

In figure~\ref{fig5} we show $m(t)$, the magnetization as a function
of $t$.  The data show no meaningful  finite size dependence, and  are in rough agreement with a $m(t)\propto t^{-5/4}$
behavior. They show however some bending that makes the precise
determination of the exponent $\delta_m$ ambiguous. Data for $e(t)$
and $q(t)$ are in similar agreement with the expected exponents
$\delta_e=\delta_q=1$.
 
\begin{figure}[htb]
\centering \includegraphics*[height=8cm,angle=270]{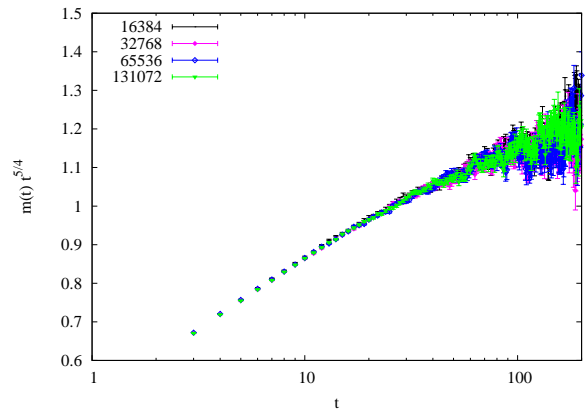} 
\caption{Ratio of $m(t)$ as in Fig.~\ref{fig5} divided by the expected
  $t^{-5/4}$ behavior, as a function of $t$.}
\label{fig6}
\end{figure}

In figure~\ref{fig6} we emphasize the deviations of the behavior of
$m(t)$ from the expected law by plotting $m(t) t^{5/4}$ as a
function of $t$ for our largest systems. Note that the ordinate range
was between $10^{-4}$ and $1$ in Fig.~\ref{fig5} and is now between
$0.6$ and $1.5$ in Fig.~\ref{fig6}.  The
ratio clearly depends on $t$ up to $t\approx 100$, where the noise
become overwhelming.  According to~\cite{CaPaRi}, the magnetization
should scale with finite $N$ like

\begin{equation}
m(t)=N^{-5/6} \ F_m(t/N^{2/3})\ .
\end{equation}
The ratio $m(t) t^{5/4}$ in figure~\ref{fig6} shows no finite size
effect but is clearly not independent of $t$. This means that we have
sizable scaling violation, and we cannot determine the exponent from
the small $t/N^{2/3}$ behavior of $m(t)$ in a scaling plot, as is
usually done (see e.g~\cite{BiCa}).
 
\begin{figure}[htb]
\centering
\includegraphics*[height=8cm,angle=270]{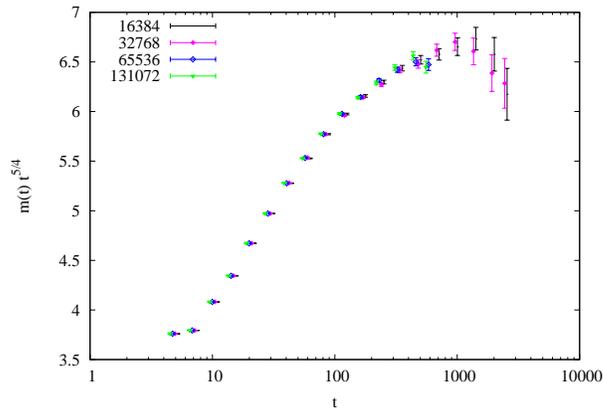}  
\caption{Magnetization $m(t)$ divided by the expected $t^{-5/4}$ behavior,
  as a function of $t$ for our largest systems. Here $m(t)$ is defined
  as the overlap between configurations at time $t$ and $t_w=3$, and
  the data for both $m(t)$ and $t$ are binned (see text). The points have
  been slightly horizontally shifted in a $N$ dependent way, in order
  to increase the clarity of the figure.}
\label{fig7}
\end{figure}

\begin{figure}[htb]
\centering
\includegraphics*[height=8cm,angle=270]{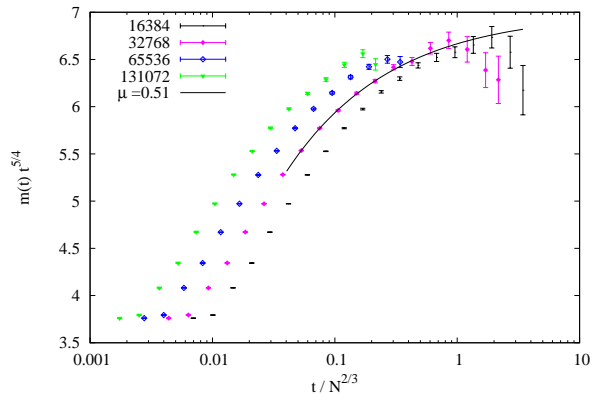}  
\caption{Scaling plot of $m(t)t^{5/4}$ (as in Fig.~\ref{fig7}) as a
  function of $t/N^{2/3}$. The data can be interpreted as showing a very slow
  $N\to \infty$ limiting constant behavior. The result of a fit of the $N=32768$  data to the form
  $m(t) t^{5/4}=a-b\big(N^{2/3}/t\bigr)^{\mu}$ is also shown. The data for the largest values of
  $t/N^{2/3}$ have a negligible weight in the fit.}
\label{fig8}
\end{figure}

To go further on, one should study the behavior of $m(t)$ for larger
values of $t$ This led us to perform additional simulations of the
largest systems $N=16384, 32768, 65536, {\rm and\ } 131072$ with two
improvements: the first is to replace~\cite{PaRaRiRu} the
magnetization, namely the overlap between spins at time $t$ and spins
at time $0$, by the overlap between spins at time $t$ and spins at
time $t_w=3$ (the precise value of $t_w$ being irrelevant),
\begin{equation}
m(t)=\frac{1}{N}\sum_i \sigma_i(t)\sigma_i(t_w)\ ,
\end{equation}
this should not change the exponents governing the large $t$ behavior,
but reduce considerably the SN ratio by increasing the signal. Next we
average the data for $m(t)$ in bins of given integer part of
$\ln(t/4)/(\ln(2)/2)$, this reduce the fluctuations and does not
affect the large $t$ behavior. We have also multiplied the number of
disorder samples by $16$ and the number of time steps by $4$ for
$N=16384, {\rm and} \ 32768$. The data can be found in Fig.~\ref{fig7},
where both $m(t)$ and $t$ are averages inside bins.  The points have
been slightly shifted (horizontally) in a $N$ dependent way, in order
to increase the clarity of the figure. Up to $t\approx 100$ there is
no visible finite size effect (even with the $16$ fold increase of the
number of disorder samples, and the corresponding decrease of the
estimated errors) and the data are steadily increasing with $t$.  We
interpret the data for larger values of $t$ as the onset of the
asymptotic $m(t)\propto t^{-5/4}$ behavior, and not as a finite size
effect, the argument being that if it was a finite size effect the data
for $N=16384$ should be below the one at $N=32768$. The question
should hopefully be settled by extending the time range to say
$t=10000$ with excellent precision up to $N=131072$. This would be
however a formidable task. Using current data it is interesting to
show a scaling plot of $m(t)t^{5/4}$ as a function of
$t/N^{2/3}$ (see Fig.~\ref{fig8}). Strong violations of scaling are apparent for small
values of the argument.  The data can be interpreted as a very slow
approach to scaling as $N$ grows, towards a limit where
$m(t)t^{5/4}$ is independent of $t$. This shows how difficult a
precise determination of $\delta_m$ is.  Even the modest task of
determining the exponent governing the first non leading term,
assuming that the leading term is exactly $\delta_m=5/4$ is difficult,
as the statistical errors on $m(t)$ increase strongly with $t$. 
Fig.~\ref{fig8} shows, as an example,
the
result of a fit of the $N=32768$ data to the form

\begin{equation}
m(t) t^{5/4}=a-b\big(N^{2/3}/t\bigr)^{\mu},
\end{equation}
in the range $t/N^{2/3}> 0.4$. Here $\mu=0.51\pm 0.01$.
The
estimates of the value of this exponent and of the plateau height
depend strongly on the range of data included in the fit however.

As a side remark we note that we made some preliminary runs using
Metropolis instead of Heat Bath and/or systematic updating instead of
random updating. It turns out that scaling violations (measured for
example as the radio of $m(t)t^{5/4}$ between $t/N^{2/3}=0.01\ {\rm
  and}\ 0.1$) are smaller with systematic updating.  It is expected
that scaling violations are not universal and can accordingly depend
significantly on the choice of a dynamics, we have no explanation
however why systematic updating gives less scaling violations than
random updating.

\section{Conclusions}

We study by Monte Carlo method the equilibrium dynamics of the one
dimensional long-range Ising spin glass model. Varying the parameter
$\sigma$ governing the decay with distance of the spin spin coupling
of this model is similar to varying the dimensionality of the
canonical Edwards-Anderson Ising spin glass model.  We extend up to
systems with $N=1024$ spins the results obtained by Monthus and Garel
in~\cite{MONGAR1D}, who argued that the dynamics is activated with
$\ln(\tau)\propto N^{\psi}$ with $\psi=1/3$ for all values of
$\sigma$, from the Sherrington-Kirkpatrick limit ($\sigma=0$) to the
physical dimension.  We find some unexpected instability in the fits,
possibly indicating critical scaling for $\tau$ (namely $\tau\propto
N^{\rho}$) at least in the non mean-field region.  It would mean that
the \oned model dynamics is similar to the EAI dynamics in the RSB
picture.

We then turn to an illustrative example of the general difficulties of
obtaining precise estimates of scaling exponents from numerical
simulations of disordered systems, by studying the decay of the
critical remanent magnetization of the Sherrington-Kirkpatrick model.
Here the exact value of the exponent is known and one can simulate
systems of very large sizes (up to $2^{17}$ here). In this example
the asymptotic regime is only approached for systems with hundred of
thousands spins, that are definitively out of reach in usual
situations, including the equilibrium dynamics of the one dimensional
long-range Ising spin glass model.


\vfill\eject

\ack Ii is a pleasure to thank Enzo Marinari and C\'ecile Monthus for
discussions, and Giorgio Parisi for raising my interest to the physics
of the remanent magnetization.  I also thank Yohan Lee-Tin-Yien for
helping me with some tricky OS problems.  The simulations where
performed on the airain and curie parallel computers in
Bruy\`eres-le-Ch\^atel (France).  Access to the later was granted by
genci (the French ``grand \'equipement national de calcul intensif")
under grant t2014056870.

\end{document}